\documentclass{aastex62}

\graphicspath{{./}{figures/}}

\received{January 1, 2018}
\revised{January 7, 2018}
\accepted{\today}
\submitjournal{ApJ}

\shorttitle{The Cyg A-2 Transient as a TDE}
\shortauthors{Tingay, Miller-Jones \& Lenc}
\begin{document}

\title{Archival VLBA observations of the Cygnus A Nuclear Radio Transient (Cyg A-2) Strengthen the Tidal Disruption Event Interpretation}

\correspondingauthor{Steven Tingay}
\email{s.tingay@curtin.edu.au}

\author[0000-0002-8195-7562]{Steven J. Tingay}
%\affil{International Centre for Radio Astronomy Research, Curtin University, Bentley, WA, Australia 6102; s.tingay@curtin.edu.au}

\author[0000-0003-3124-2814]{James C. A. Miller-Jones}
\affil{International Centre for Radio Astronomy Research, Curtin University, Bentley, WA, Australia 6102}

\author{Emil Lenc}
\affil{CSIRO Astronomy and Space Science, P.O. Box 76, Epping, NSW 1710, Australia}

%% Note that the \and command from previous versions of AASTeX is now
%% depreciated in this version as it is no longer necessary. AASTeX 
%% automatically takes care of all commas and "and"s between authors names.

%% AASTeX 6.2 has the new \collaboration and \nocollaboration commands to
%% provide the collaboration status of a group of authors. These commands 
%% can be used either before or after the list of corresponding authors. The
%% argument for \collaboration is the collaboration identifier. Authors are
%% encouraged to surround collaboration identifiers with ()s. The 
%% \nocollaboration command takes no argument and exists to indicate that
%% the nearby authors are not part of surrounding collaborations.

%% Mark off the abstract in the ``abstract'' environment. 
\begin{abstract}
We have analyzed archival VLBA data for Cygnus A between 2002 and 2013, to search for radio emission from the transient discovered in 2015 by \citet{per18} approximately 0.4\arcsec~ from the nucleus of Cygnus A (Cyg A-2).  \citet{per18} use VLA and VLBA archival data (between 1989 and 1997) to show that the transient rises in flux density by a factor of at least five in less than approximately 20 years.  With the additional data presented here, we revise the rise time to  between approximately four years and six years, based on a new detection of the source at 15.4 GHz from October 2011.  Our results strengthen the interpretation of Cyg A-2 as the result of a Tidal Disruption Event (TDE), as we can identify the location of the compact object responsible for the TDE and can estimate the angular expansion speed of the resulting radio emitting structures, equivalent to an apparent expansion speed of $<0.9c$. While our results are consistent with recent X-ray analyses, we can rule out a previously suggested date of early 2013 for the timing of the TDE.  We favour a timing between early 2009 and late 2011.  Applying the model of \citet{nak11}, we suggest a TDE causing a mildly relativistic outflow with a (density-dependent) total energy $>10^{49}$ erg.  Due to the improved temporal coverage of our archival measurements, we find that it is unlikely that Cyg A-2 has previously been in a high luminosity radio state over the last 30 years. 

\end{abstract}

%% Keywords should appear after the \end{abstract} command. 
%% See the online documentation for the full list of available subject
%% keywords and the rules for their use.
\keywords{galaxies: individual (Cygnus A) - radiation mechanisms: general - techniques: high angular resolution - radio continuum: general}

%% From the front matter, we move on to the body of the paper.
%% Sections are demarcated by \section and \subsection, respectively.
%% Observe the use of the LaTeX \label
%% command after the \subsection to give a symbolic KEY to the
%% subsection for cross-referencing in a \ref command.
%% You can use LaTeX's \ref and \label commands to keep track of
%% cross-references to sections, equations, tables, and figures.
%% That way, if you change the order of any elements, LaTeX will
%% automatically renumber them.
%%
%% We recommend that authors also use the natbib \citep
%% and \citet commands to identify citations.  The citations are
%% tied to the reference list via symbolic KEYs. The KEY corresponds
%% to the KEY in the \bibitem in the reference list below. 

\section{Introduction} \label{sec:intro}

Ushered in by a new era of data processing capacity and high survey speed telescopes, significant effort has been expended over the past decade into large-scale surveys for radio transients at centimetre to metre wavelengths, and for transient durations longer than a second, e.g. \cite{ku20,hej19,bell19,fen17,mur13}.

These surveys have, in general, revealed remarkably few transient radio objects in a systematic manner, as elaborated by \cite{met15}.  It is likely that significantly higher sensitivity is required to uncover more of the slow transient and variable radio universe, for example using the Square Kilometre Array \citep{fen15}.  However, some significant recent advances have been forthcoming from targeted follow-up observations of interesting events, for example the multi-wavelength and multi-messenger observations of GW170817 \citep{abb17}.

It is still the case that many interesting radio transients are the result of serendipity, rather than surveys.  One such recent discovery, by virtue of the fact that the transient resides within the host galaxy of one of the most powerful and nearby radio galaxies, Cygnus A, was reported by \citet{per18}.  

Cygnus A has been observed with high angular resolution radio telescopes over decades, providing a high quality and long timescale series of observations.  So when VLA observations in 2015 revealed a never-before-seen compact radio source approximately 0.4\arcsec\ from the bright nucleus of Cygnus A, \citet{per18} could examine historical data.  They found that no such source had existed in the 1980s and 1990s, from VLA and VLBA data.

\citet{per18} made new observations in the frequency range 7.1 -- 47 GHz with the VLA and VLBA to examine the object, designated Cygnus A-2 (Cyg A-2), and from the sum total of the available data favoured a scenario in which Cyg A-2 resulted from accretion onto a secondary supermassive black hole, approximately 450 pc distant from the primary black hole forming the Cygnus A nucleus.  The rapid rise in flux density, by a factor of at least five in approximately 20 years, was cited as one of the distinctive characteristics of this object and evidence in favour of an explanation in terms of accretion.  \citet{per18} note, in particular, the possibility that Cyg A-2 is the result of a Tidal Disruption Event (TDE).

Additional suggestions of a TDE origin have been presented by \citet{dev19}, from an analysis of new and historical X-ray observations of Cygnus A.  They find an enhancement in the X-ray light curve in early 2013 and note previous results that indicate the presence of a short-lived, fast, ionised outflow near that epoch.  \citet{dev19} take this as evidence for a TDE at that time, with the radio emission seen by \citet{per18} in 2015 due to an afterglow, although they do not rule out stochastic X-ray variation of the Cygnus A AGN, which is not resolved from Cyg A-2.

We examine the TDE origin for Cyg A-2 in more detail, by the additional mining of archival data to constrain more tightly the rise time of the radio emission.  We have extracted archival VLBA data between 2002 and 2013, over frequencies ranging from 1.6 GHz to 15.4 GHz, and search for emission from the reported location of Cyg A-2.  We detect Cyg A-2 at one epoch, at a frequency of 15.4 GHz, in late 2011.  This detection has consequences for the interpretation of the other available data for Cyg A-2 in the context of a TDE explanation.  In \S 2 we describe the archival data and the results of our data processing.  In \S 3 we discuss our results in light of the suggestion by \citet{per18} and \citet{dev19} that Cyg A-2 is the result of a TDE.

\section{Data Processing and Results}

All data were extracted from the VLBA archive, following a search for data corresponding to the position of Cygnus A, with the dates (column 1) and project codes (column 2) for the observations listed in Table \ref{tab:table1}.  The data for projects BL1788 and BP171 data were correlated with DiFX \citep{del11,del07}.  All other data were correlated using the original VLBA hardware correlator.  The central frequencies (column 3), bandwidths (column 4), channels per sub-band (column 5), and observation durations (column 6) are also given in Table \ref{tab:table1}.

The correlated data were imported into AIPS \citep{van96} for standard processing via fringe-fitting and calibration of the visibility amplitudes.  The partially calibrated data were then exported from AIPS and imported into DIFMAP \citep{she94} for imaging and self-calibration.  The imaging process initially concentrated on the bright core and milliarcsecond-scale jet in Cygnus A, utilising this bright structure to self-calibrate the visibilities in amplitude and phase.  For each observation, the expected core-jet structure was recovered, appropriately for the angular resolution (column 7 of Table \ref{tab:table1}) and sensitivity (column 9 of Table \ref{tab:table1}) of each observation.

Once the data were well calibrated using the bright core-jet, a wider field of view image could be formed in order to search for emission at the location of the transient: $\alpha=$19:59:28.32345; $\delta=+$40:44:01.9133 (J2000, $\pm$1 mas), as reported in \citet{per18}.  In this case, care needs to be taken of temporal and bandwidth smearing.  In all but two cases, for the data for BP171 and BL178, the frequencies, IF bandwidths, time averaging, and maximum baseline lengths were such that the expected maximum signal loss due to smearing is less than 5\%.  For these observations, the IF bandwidths produced higher signal loss due to smearing, so these observations were imaged using individual channels, greatly reducing the smearing.  The maximum losses due to smearing are listed in Table \ref{tab:table1} (column 10).

\begin{figure}
\epsscale{0.8}
\plotone{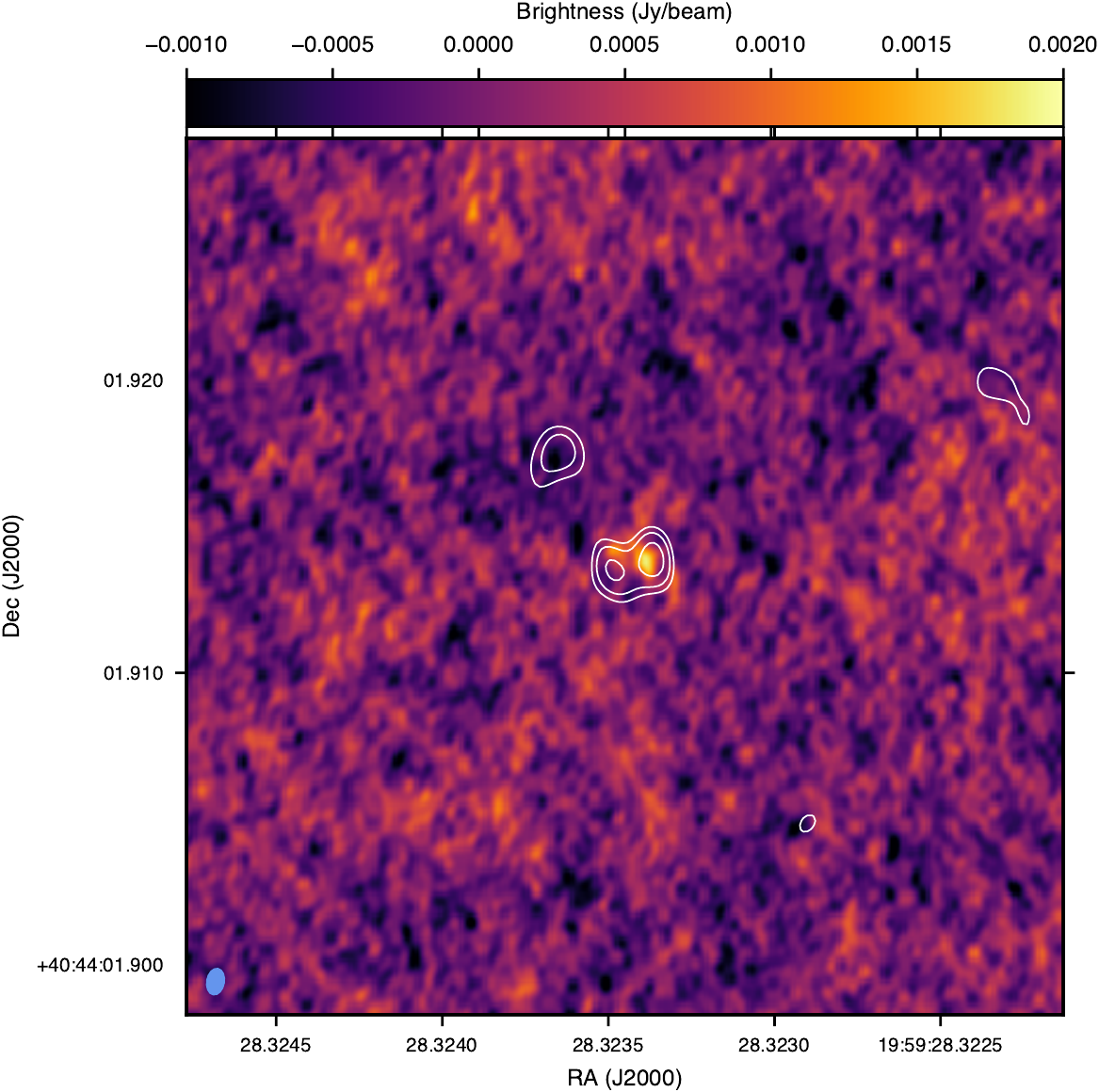}
\caption{VLBA images of Cyg A-2.  Color scale shows the 15.4\,GHz image from 2011 October 3, with the beam size shown as the blue ellipse at the bottom left.  Contours show the stacked 8.4-GHz image from three observations taken in 2016 November \citep[re-processing of the BP213 data presented by][]{per18}.  Contours are at levels of $\pm0.6(\sqrt{2})^n$\,mJy\,beam$^{-1}$, where $n=1,2,3...$.}\label{fig:f0}
\end{figure}

We detected Cyg A-2 at one epoch, the 15.4 GHz observation in October 2011 (BL178; Figure \ref{fig:f0}), at a level of approximately 7$\sigma$ (column 8 of Table \ref{tab:table1}).  All other epochs processed resulted in non-detections.  For the non-detections, the image RMS values vary with frequency, duration of observation, and overall quality of the data, as also listed in Table \ref{tab:table1} (column 8).

\begin{deluxetable}{cccccccccc}[h]
\tablecaption{Archival VLBA data for Cyg A-2.  Columns indicate the observation date, VLBA project code, central frequency ($\nu_{c}$), number and bandwidth (in MHz) of each IF ($\Delta \nu$), number of channels per IF ($N_{chan}$), time on source ($\Delta T$), the major and minor axes and position angle of the synthesized beam ($\theta_{max}$/$\theta_{min}$/PA) , the peak intensity of detections, the RMS noise level at the position of Cyg-A-2, and the expected signal loss due to bandwidth smearing ($L$).  \label{tab:table1}}
\tablehead{
\colhead{Date} & 
\colhead{Proj.} & 
\colhead{$\nu_{c}$} & 
\colhead{$\Delta \nu$} & 
\colhead{$N_{chan}$} & 
\colhead{$\Delta T$} &
\colhead{$\theta_{max}$/$\theta_{min}$/PA} & 
\colhead{$S_{peak}$} &
\colhead{RMS} &
\colhead{L} \\
\colhead{(yyyy.mm.dd)} & 
\colhead{} & 
\colhead{(GHz)} &
\colhead{(MHz)} &
\colhead{} & 
\colhead{(min)} & 
\colhead{(mas/mas/$^{\circ}$)} &
\colhead{(mJy/beam)} &
\colhead{(mJy/beam)} &
\colhead{(\%)} \\
}
\startdata
2002.08.06&BC123&1.550&8$\times$8&256&14&15.1/9.4/$-$20&-&2&2.0 \\
2006.06.15&BL137&8.268&4$\times$8&64&7&1.3/1.0/50&-&4&4.7 \\
2009.02.08&BC185&4.845&4$\times$8&64&82&3.3/2.3/$-$23&-&0.6&1.2 \\
%2010.06.04&BL1788&15.357&4$\times$8&64&37&0.8/0.5/$-$12&0.4&4.7 \\
2011.10.03&BL178&15.357&8$\times$8&32&37&0.8/0.5/$-$10&2.3 & 0.3 &0.04 \\
2013.07.17&BP171&4.360&8$\times$32&4096&7&5.9/1.8/$-$22&-&2&0.04\\
\enddata
\end{deluxetable}

In general, the observations are not of long duration and do not contain the full set of VLBA antennas.  However, all datasets contain adequate information to produce good images.  Other data exist in the VLBA archive, but we found them to be of low quality. The best limits and the detection are obtained from the longest-duration observations.  We have included data from BC123 at 1.6 GHz, as it produces a good image.  However, the reader is cautioned regarding this particularly low frequency as the scattering effects in this region of Cygnus A are significant \citep{per18}.

Finally, in order to verify our data processing methods, and provide confidence in our detection and non-detections, we downloaded the VLBA data presented by \citet{per18}, with project code BP213. We found that missing antennas on 2016 November 14 led to a higher noise level for that epoch, causing us to omit it from the stacked image. Owing to the proximity of Cyg A-2 to the Cygnus A AGN radio source, we self-calibrated the data on Cygnus A, and simultaneously imaged both fields within AIPS.  Thus our complex gain solutions were derived for the target data themselves, rather than being interpolated to Cyg A-2 from neighbouring scans on Cygnus A, as was the case in \citet{per18}. Applying these methods, we accurately reproduced the results presented in \citet{per18}. Figure \ref{fig:f0} shows our image from October 2011, with contours from our image of the \citet{per18} data from 2016 November.

From Figure \ref{fig:f0}, we see that at 15.4 GHz in late 2011, we detect one of the possible two components detected approximately four years later by \citet{per18}.  To align the two images, we had to correct for both the difference in the positions of Cygnus A assumed at correlation (a shift of 0.53\,mas along a position angle 309\degr\ E of N), as well as the known frequency-dependent core shift of Cygnus A \citep[0.49\,mas along a position angle of 284\degr;][]{Bach04,Nak19}.  Having done this, we find that the 15.4-GHz detection from 2011 aligns with the western lobe of the 8.4-GHz image from November 2016.  \citet{per18} note the possibility that the extension seen in 2016 is due to phase referencing errors. However, we recovered the same structure by self-calibrating the data using the Cygnus A radio AGN in the same set of visibilities, negating the need for any phase transfer.  We therefore suggest that the extended structure is likely to be real, and take our 2011 detection to represent the early time, unresolved radio emitting structure, presumably marking the origin of a jet-like feature that subsequently evolved and was then detected by \citet{per18}.  If the same jet-like structure seen in 2016 was present in late 2011, our image would have the sensitivity to detect this extension.

Under this interpretation, we can estimate the angular expansion speed for the jet-like feature.  From Figure \ref{fig:f0} we measure the source extension to be $1.4\pm0.1$ mas ($\sim$1.3 pc at the distance of Cygnus A). Were the expansion to have begun at the time of the 15.4-GHz observation, the inferred angular expansion rate over the intervening five-year period would be 0.25\,mas\,yr$^{-1}$, corresponding to a rest-frame apparent expansion speed of $0.9c$.  However, recognising that the zero separation time likely occurred prior to the 15.4 GHz observation, we treat this value as an upper limit.
%Over the course of the five years between observations the angular expansion rate is therefore $\sim$0.3 mas/yr, corresponding to an apparent expansion speed of $<1.0c$, recognising that the zero separation time likely occurred prior to the 15.4 GHz observation.

\section{Discussion and Conclusion}

\begin{figure}
\plotone{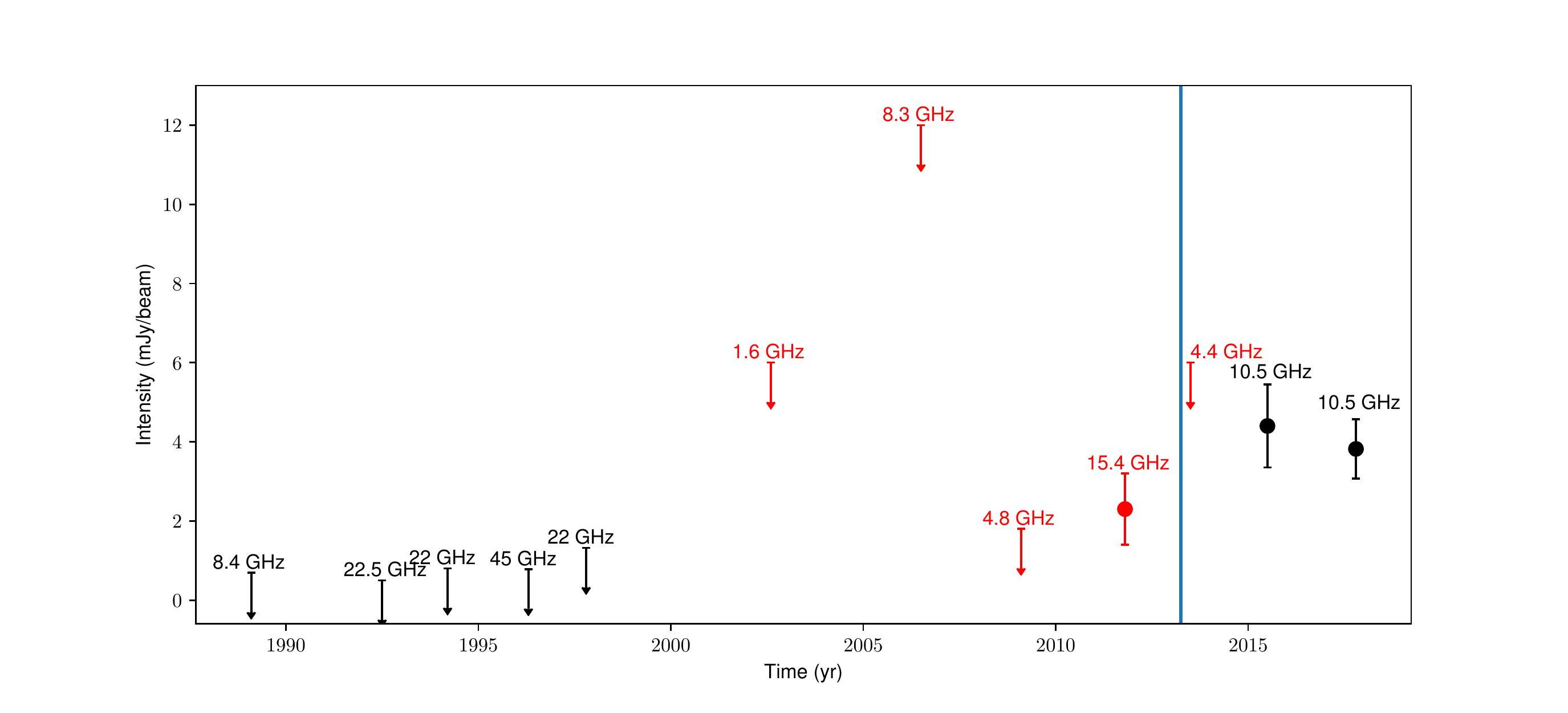}
\caption{Radio light curve compiled from published results of \citet{per18} (black upper limits and measured points) and archival data processed and presented here (red upper limits and detection).  Rather than plot all the measurements from \citet{per18} from 2015 - 2017, we plot representative points in 2015 and 2017 at the common frequency of 10.5 GHz.  The vertical blue line marks the February/March 2013 epoch of the 2 - 10 keV peak noted by \cite{dev19} as their proposed timing of the TDE.  Upper limits and error bars are all 3$\sigma$.\label{fig:f1}}
\end{figure}

Figure \ref{fig:f1} summarises our results in the context of the results published by \citet{per18}, showing that comparable upper limits at comparable frequencies now extend from 1989 to 2010, relative to the 2015 discovery observations of Cyg A-2. Consistent with the upper limits presented in \citet{per18}, our upper limits (and errors on the 15.4 GHz detection from 2011) are three times the measured RMS values from our images at the location of Cyg A-2\footnote{\citet{per18} do not list the significance of their upper limits, but we clarified them to be 3$\sigma$ via private communication with the authors.}.  We note that our upper limits span a range of 1.6 -- 8.3 GHz and that the spectral behaviour of Cyg A-2 presented by \citet{per18} could possibly contribute to non-detection at the lower frequencies, 1.6 -- 4.8 GHz.

The non-detections, plus our detection at 15.4 GHz in late 2011, immediately constrain the rise time for Cyg A-2 to a period significantly shorter than the approximate 20-year period presented in \citet{per18}. We define the rise time as the period between the TDE and the peak flux density observed by \citet{per18} in mid 2015.  Our October 2011 detection with the VLBA places a lower limit on the rise time of $\sim 4$ years, which given the unresolved nature of the source in that observation, is likely to be relatively close to the actual time of the TDE. The February 2009 VLBA non-detection then defines a $\sim$6 year upper limit on the rise time, albeit with the caveat that it was taken at a relatively low frequency, which may be below the self-absorption frequency. 
Thus, with the caveats and definitions outlined above, we constrain the rise time to a range of $\sim$4 to $\sim$6 years.
  
% which we take as in the range February 2009 and October 2011, 
  
% However, it is worth noting that   With these caveats and definitions, we adopt a $\sim$4 year to $\sim$6 year rise time.}

As well as constraining the rise time, a more complete sampling of archival data over the last almost 30 years also provides some evidence that the appearance of Cyg A-2 is likely not a recurring event.

The conclusions of \citet{per18} regarding a possible supernova origin are unchanged by these new data.  The stricter constraint on rise time does not provide strong additional support for or against a supernova origin, as the classes of supernova discussed by \citet{per18} all have rise times significantly shorter than five years.  In the context of our estimated 4 $-$ 6 year limit on the rise time, the four year lower limit poses a challenge to these models, but does not rule them out.  However, the challenges for supernova models based on energetics described in \citet{per18} remain.

In terms of scenarios involving accretion onto a secondary black hole, radio emission from a relatively steady state accretion is well known to be associated with AGN and stellar mass black holes in our Galaxy.  Powerful, relativistically beamed jets in AGN regularly display flares with rise times consistent with the rise time limits and flux density increase factors for Cyg A-2 \citep{par16}.  Generally many such flares (often overlapping in time) are seen in multi-decade periods.  The apparent luminosity of Cyg A-2 being many orders of magnitude below the apparent luminosities of AGN jets and the singular appearance of Cygnus A-2  in a $\sim$30 year period do not argue in favour of this scenario,  although a lower limit of four years on the rise time may admit AGN flare timescales; significantly shorter rise times than this would be more difficult for AGN models to achieve.

Galactic microquasars such as GRS 1915$+$105 \citep{mir94} and GRO J1655$-$40 \citep{tin95} have also been show to have relativistic jets and, in contrast to AGN, often remain dormant for long periods of time (decades).  However, if such an object were placed at the distance of Cygnus A and directed at us, unrealistically large Doppler boosting factors are required to produce a radio source as bright as the few mJy seen for Cyg A-2.

Thus, the current evidence points to a luminosity consistent with accretion onto a massive black hole, but that is not steady state or episodic and frequent.  This favours a discrete accretion event, such as a TDE, an option noted by \citet{per18} based on the presence of an optical/NIR counterpart assumed to be persistent. Our results provide additional weight to this scenario by showing that the near-infrared counterpart observed in 2002 May \citep{can03} existed prior to the radio transient and so was not itself a transient. As noted by \citet{per18}, an explanation of the near infrared counterpart as either steady emission from a secondary active galactic nucleus \citep{per18} or the stripped core of a merging galaxy \citep{can03} would be consistent with a TDE scenario for the radio transient. Ongoing monitoring of the infrared counterpart to Cyg A-2 could assist in confirming this suggestion.

A handful of TDEs and candidate TDEs have been shown to produce radio emission, consistently showing radio light curves with rise times of months following high energy flares, as comprehensively reviewed by \cite{de19} and more recently by \cite{alexander20}.  An analysis of approximately 31 years of X-ray data for Cygnus A by \citet{dev19} shows a mild enhancement in the 2 - 10 keV luminosity in February/March 2013; the authors also note previously published evidence for a short-lived, fast, ionised outflow in the NuSTAR spectrum near the same epoch.  \citet{dev19} suggest that Cyg A-2 is the afterglow of a TDE marked by this enhanced X-ray emission.

The timing of the X-ray enhancement is shown in Figure \ref{fig:f1}, in relation to the radio measurements of \cite{per18} and our new VLBA archival measurements.  Our new detection at 15.4 GHz prior to this suggested 2013 timing of the TDE clearly rules out this suggestion.  \citet{dev19} do not rule out that the 2013 X-ray enhancement can be explained as the stochastic variation of the Cygnus A AGN.  Our result is consistent with this interpretation.  However, jetted TDEs with X-ray emission such as SWIFT J1644$+$57 show a rapid rise in X-rays, variation in a high state over 100s of days, and sharp/gradual declines over 1000s of days \citep[e.g.][]{lev16}.  Other similar examples are known \citep{kom15}.  Also, the ionised outflow noted by \citet{dev19} could have appeared at any point between 2005 and 2013. 
The  X-ray lightcurve of \citet{dev19} is sparse, with measurements from both {\it NuSTAR} and {\it Swift} contributing to the identification of the X-ray enhancement in early 2013, but the closest measurement previous to that was in late 2008.  \citet{dev19} also note the uncertainties on the errors in the X-ray luminosities when noting the statistically significant X-ray enhancement in 2013.  We suggest that, while a 2013 TDE date is ruled out by our new VLBA detection, the X-ray enhancement seen in 2013 may be the result of a TDE approximately two to four years earlier, but representing a decaying X-ray emission rather than marking the timing of the TDE itself.  \citet{dev19} place the observed X-ray enhancement at approximately $2\times10^{44}$ erg\,s$^{-1}$ above the long term value for Cygnus A, which appears plausible for an X-ray afterglow of a jetted TDE on a timescale of years post-disruption; the jetted TDEs Swift J1644$+$57 and Swift J2058$+$05 maintained an X-ray luminosity above $10^{44}$ erg\,s$^{-1}$ for more than a year post-disruption \citep{lev16,auch17}.

\begin{figure}
\plottwo{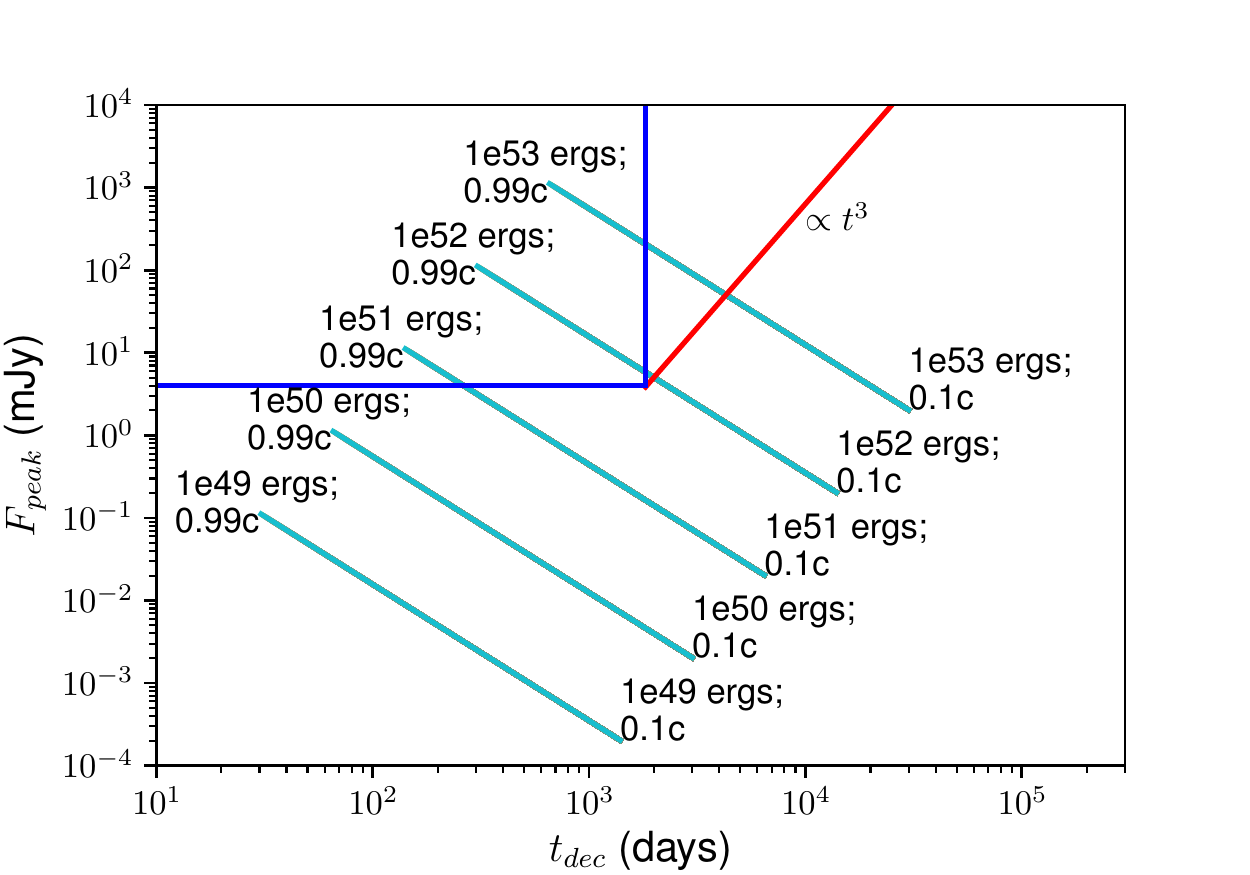}{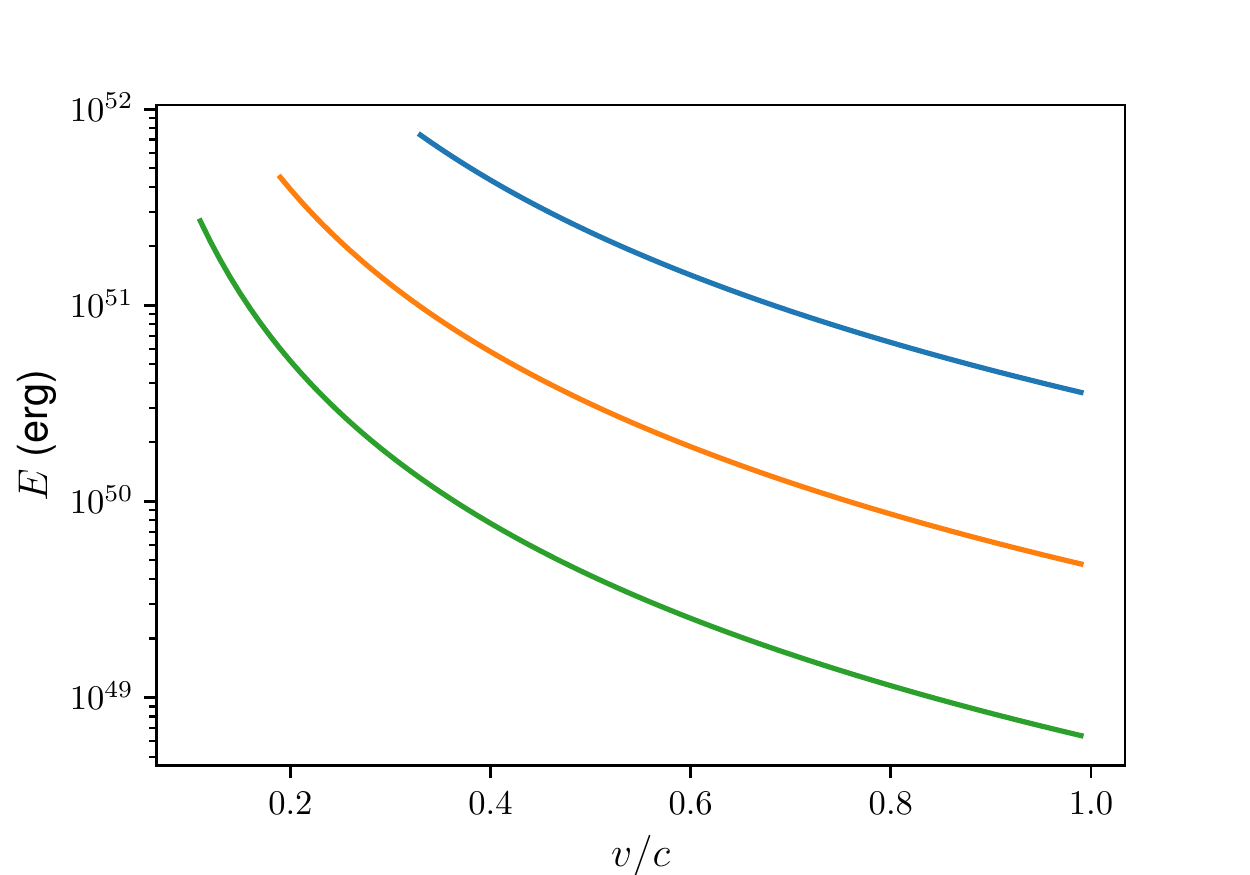}
\caption{Left panel: The peak flux density, $F_{\rm peak}$, as a function of time to reach the peak, $t_{\rm dec}$, for a range of values of outflow energy, $E$, and outflow speed, $\beta$, as described in the text, for $n=1$\,cm$^{-3}$.  The dark blue lines indicate constraints from the observed flux density of 4 mJy and the limit on the observed rise time, {\bf 5 years}, indicating the family of models that produce peak flux densities of plausible magnitude and timescale.  The red diagonal line indicates the pre-peak flux density evolution proportional to $t^{3}$, such that values of $E$ and $\beta$ above that line produce a pre-peak flux density greater than 4 mJy, within {\bf 5 years} of the TDE.  Right panel: all values of $E$ and $\beta$ that produce a peak flux density above 4 mJy within {\bf 5 years} of the TDE lie above the line.  In this case, the blue line represents the nominal model we have adopted, with $n=1$\,cm$^{-3}$.  The orange line represents the model with $n=10$\,cm$^{-3}$, and the green line represents the model with $n=100$\,cm$^{-3}$, as described in the text.} \label{fig:f2}
\end{figure}

Our estimate of the apparent expansion speed of the radio structure, originating from an apparently fixed point we take to be the location of the black hole and TDE, implies a mildly relativistic intrinsic expansion.  The interpretation as a jetted TDE is consistent with our identification of a location of origin and a jet-like evolution of the structure.  The radio structure is one-sided and has an apparent expansion speed of $<0.9c$, implying a relatively small angle to the line of sight for the jet, although the true apparent speed is unlikely to be as high as $c$.

Based on this scenario, and as typically adopted for TDEs \citep{svv16,sto16}, we use a canonical blast wave model from a mildly-relativistic outflow \citep{nak11} to explore the energetics and timescales of Cyg A-2. While the jet kinetic power is expected to evolve on timescales of days \citep{mgm12}, at late times the blast wave (producing the radio emission) is likely to become mildly relativstic and spherical \citep[e.g.][]{sto16}. In this model, the late-time synchrotron emission from a spherically expanding outflow at frequency $\nu_{\rm obs}$ peaks at a time $t_{\rm dec}$ (following the X-ray flare) at a peak flux density of $F_{\rm peak}$, as a function of the total energy injected into the outflow, $E$, and the outflow speed, $\beta=v/c$.  

% Based on this scenario, we use the model of \citet{nak11} to explore the energetics and timescales of Cyg A-2 based on a mildly relativistic outflow to represent the TDE.  

We adopt the canonical values of \citet{nak11} for the density of the environment into which the outflow expands ($n=1 {\rm \,cm}^{-3}$) and the fraction of the total energy carried by electrons and magnetic fields ($\epsilon_{B}=\epsilon_{e}=0.1$).  \citet{nak11} cite the power-law index for the electron population, $p$, in the range $2 - 3$.  Here we adopt $p=2.5$.  Before the emission reaches $F_{peak}$ at $t_{dec}$, the emission rises proportionally to $t^{3}$.  In recognition of the uncertainty in the value of $n$ that we adopt (the comprehensive study of the infrared counterpart by \citealt{can03} is silent on the density of its environment), we also note results for $n=10 {\rm \,cm}^{-3}$ and $n=100 {\rm \,cm}^{-3}$ \citep[encompassing the range of densities inferred for the environment of the TDE XMMSL1 0740$-$85;][]{ale17}, to demonstrate their sensitivity to $n$.  We also explore models that assume rise times of 4, 5, and 6 years, in order to cover our observational estimates.

%n=1   >0.36 >6e51 0.99) >4e50
%n=10  >0.21 >3e51 0.99 >5e49
%n=100 >0.12 >2e51 0.99 >6e48

The left panel of Figure \ref{fig:f2} shows how $F_{peak}$ varies as a function of $t_{dec}$, for $n=1$\,cm$^{-3}$ and a range of values of $E$ and $\beta$, relative to {\bf a} rise time for the TDE of 5 years (1825 days), representing the mid-point of our 4 $-$ 6 year rise time range estimated from Figure \ref{fig:f1} and the detection level of $\sim$4 mJy, at a frequency of $\nu_{obs}=10$ GHz.  The figure shows that a range in $E$ and $\beta$ values can produce an appropriate peak flux density on the correct timescales for Cyg A-2.  The diagonal line on the figure labeling the $t^{3}$ rise for the pre-peak flux density denotes values of $E$ and $\beta$ above which the flux density reaches 4 mJy prior to peak flux density, but within {\bf 5 years} of the TDE.  All of these solutions are valid in the framework of \citet{nak11}, as $\nu_{obs}=10$ GHz is greater than both the characteristic synchrotron frequency for the typical electron Lorentz factor ($\leq1$ GHz for these solutions) and the characteristic synchrotron self-absorption frequency ($\leq3$ GHz for these solutions).

All values of $E$ and $\beta$ that produce a flux density greater than 4 mJy within {\bf 5 years} of the TDE event are reflected as the regions above the lines in the right panel of Figure \ref{fig:f2}, for our full range of assumed densities ($n=1$, $10$ and $100$\,cm$^{-3}$).  
%Over the range of allowed values of $\beta$ ($>0.36$ for $n=1$\,cm$^{-3}$), all values of $E$ above the line produce flux densities above 4 mJy within 1400 days of the TDE event.  
The minimum $\beta$ required to accommodate the observational constraints drops from 0.36 to  0.10 as the density increases from 1 to 100\,cm$^{-3}$ over the rise time range.  Lower values of $\beta$ and lower densities both require higher energies; as $\beta$ decreases, the minimum required energy increases from $\sim4\times10^{50}$ to $\sim6\times10^{51}$\,erg for $n=1$\,cm$^{-3}$ and a rise time of 4 years, whereas for $n=100$\,cm$^{-3}$, the minimum required energy goes from $\sim6\times10^{48}$ to $\sim9\times10^{51}$\,erg for a rise time of 6 years.
%At high values of $\beta$, energies above $\sim4\times10^{50}$ erg are required for $n=1$\,cm$^{-3}$, and at the bottom end of the allowed range in $\beta$, energies greater than $\sim6\times10^{51}$ erg are required.  With $n=10 {\rm \,cm}^{-3}$, the allowed range of $\beta$ is $>0.21$, with $E>3\times10^{51}$ erg at low $\beta$ and $E>5\times10^{49}$ erg at high $\beta$.  With $n=100 {\rm \,cm}^{-3}$, the allowed range of $\beta$ is $>0.12$, with $E>2\times10^{51}$ erg at low $\beta$ and $E>6\times10^{48}$ erg at high $\beta$.  
From an outflow energetics point of view, these numbers appear reasonable, as typical outflow kinetic energies for TDEs range from $10^{48}$\,erg for thermal TDEs to several times $10^{51}$\,erg for the non-thermal TDE Sw J1644+57 \citep{alexander20}, and can easily be accommodated by the expected energy release for a TDE that involves half a stellar mass \citep[$10^{52} - 10^{53}$\,erg;][]{lu18}.  This is also consistent with the TDE bolometric energies and efficiencies of rest mass conversion to energy compiled by \citet{moc20}, which span ranges of approximately $10^{50}-10^{52}$ erg and $0.001 - 0.1$, respectively.

In summary, our new archival VLBA measurements of Cyg A-2, along with the radio data from \cite{per18}, the analysis of \cite{dev19}, and consideration of a simple model for the radio emission in temporal and energetic terms therefore strengthen the general interpretation of a TDE origin for Cyg A-2.

\acknowledgements
We thank the anonymous referee for providing comments that led to significant improvements to the paper, by prompting us down a route of analysis that yielded important new results (the detection at 15.4 GHz).  This research was partially supported by the Australian Government through the Australian Research Council's Discovery Projects funding scheme (project DP200102471).  This work made use of the Swinburne University of Technology software correlator, developed as part of the Australian Major National Research Facilities Programme and operated under licence.  This research has made use of NASA's Astrophysics Data System.  AIPS is produced and maintained by the National Radio Astronomy
Observatory, a facility of the National Science Foundation
operated under cooperative agreement by Associated Universities, Inc.

\facility{VLBA}

%% This command is needed to show the entire author+affilation list when
%% the collaboration and author truncation commands are used.  It has to
%% go at the end of the manuscript.
%\allauthors

%% Include this line if you are using the \added, \replaced, \deleted
%% commands to see a summary list of all changes at the end of the article.
%\listofchanges

\end{document}